\documentclass[aps,twocolumn,english,aip,superscriptaddress,amssymb,showpacs,preprintnumbers,amsmath,amssymb,superscriptaddress]{revtex4-1}
\usepackage[utf8]{inputenc}
\usepackage{amsmath}
\usepackage{graphicx}
\usepackage{dcolumn}
\usepackage{bm}
\usepackage{nicefrac}
\usepackage{color}

\newcommand{\nc}{\newcommand}
\nc{\be}{\begin{equation}}
\nc{\ee}{\end{equation}}
\nc{\bea}{\begin{eqnarray}}
\nc{\eea}{\end{eqnarray}}
\nc{\bean}{\begin{eqnarray*}}
\nc{\eean}{\end{eqnarray*}}
\nc{\mb}{\mbox}
\nc{\vp}{\mb{\bf p}}
\nc{\vn}{\mb{\bf n}}
\nc{\vq}{\mb{\bf q}}
\nc{\rr}{\mb{\bf r}}
\nc{\vz}{\hat {\mb{\bf z}}}
\nc{\vj}{\mb{\boldmath$j$}}
\nc{\vg}{\mb{\boldmath$g$}}
\nc{\x}{\mb{\boldmath$x$}}
\nc{\A}{\mb{\boldmath$A$}}
\nc{\va}{\mb{\boldmath$a$}}
\nc{\vs}{\mb{\boldmath$\sigma$}}
\nc{\vpi}{\mb{\boldmath$\pi$}}
\nc{\nab}{\nabla}
\nc{\X}{\sf x}
\nc{\ep}{\epsilon}

\begin{document}

\title{Gate Controlled Majorana Zero Modes of a Two-Dimensional Topological Superconductor}

\author{Nima Djavid}
\affiliation{Department of Electrical engineering, University of California Riverside,
Riverside, California 92521, USA}

\author{Gen Yin}
\affiliation{Department of Electrical engineering, University of California at
Los Angeles, Los Angeles, California 90095, USA}

\author{Yafis Barlas}
\affiliation{Department of Electrical engineering, University of California Riverside,
Riverside 92521, USA}

\author{Roger K. Lake}
\thanks{rlake@ece.ucr.edu}
\affiliation{Department of Electrical engineering, University of California Riverside,
Riverside 92521, USA}

\begin{abstract}
Half-integer conductance, the signature of Majorana edge
modes, was recently observed in a thin-film
magnetic topological insulator / superconductor bilayer.
This letter analyzes a scheme for gate control
of Majorana zero modes in such systems.
Gating the top surface of the thin-film magnetic topological insulator controls the
topological phase in the region underneath the gate.
The voltage of the transition depends on the gate width, and narrower gates require larger voltages.
Relatively long gates are required, on the order of 2 $\mu$m, to prevent hybridization of the end modes
and to allow the creation of Majorana zero modes at low gate voltages.
Applying voltage to T-shaped and I-shaped gates localizes the Majorana zero modes at their ends.
This scheme may provide a facile method for implementing quantum gates for topological quantum computing.
\end{abstract}
\maketitle

\section{Introduction}
Majorana fermions are charge-neutral fermionic particles that are their own antiparticles
originally proposed by Ettore Majorana \cite{majorana_symmetric_2006}.
Prior theoretical work suggested that Majorana fermions
could exist in topological superconductors as elementary
excitations \cite{kitaev_unpaired_2001,fu_superconducting_2008,Wimmer_PRL2010,Elliott_Franz_RMP15}.
The first experimental demonstration was the zero-bias anomaly observed
in a III-V semiconductor nanowire
coupled to an s-wave superconductor \cite{mourik_signatures_2012,deng_majorana_2016},
a material system that is a physical implementation of
Kitaev's one dimensional topological superconductor model \cite{kitaev_unpaired_2001}.
In the middle of the superconducting gap, zero-energy localized states appear at the
ends of such nanowires.
These states are Majorana zero modes (MZMs).
MZMs obey non-Abelian statistics \cite{alicea_non-abelian_2011},
and they can be used for fault-tolerant, topological
quantum computing \cite{kitaev_fault-tolerant_2003,alicea_new_2012}.
This requires the precise control of the position of the MZMs in a nanowire network.
Gate control has been proposed and analyzed
\cite{alicea_new_2012,aasen_milestones_2016},
and recent experiments demonstrated prototypes of such nanowire
networks \cite{plissard_formation_2013}.

Although previous experiments using III-V nano-wires have shown exciting possibilities,
an implementation in two-dimensional (2D) thin films would be more compatible with conventional
semiconductor device fabrication.
Such a scheme in the InAs / superconducting system has been proposed \cite{2D_InAs_SC_PRL17}.
Recently, quantized half--integer conductance ($\frac{e^{2}}{2h}$) was observed in a different
material system consisting of a thin film
magnetic topological insulator (MTI)
capped by an s--wave superconductor (Nb) \cite{he_chiral_2017}.
The half--integer conductance suggested the existance of a chiral Majorana mode propagating
along the edge\cite{he_chiral_2017,lian_edge-state-induced_2016,wang_chiral_2015,chen_effects_2017}.
In this system, the MTI consisted of Cr doped, epitaxial, thin film (Bi,Sb)$_2$Te$_3$.
Recently, signatures of MZMs were observed in a similar material system
by scanning probe measurements, where the anti-periodic boundary condition is induced by a superconducting
vortex\cite{sun_majorana_2016}.
Prior theoretical studies on the MTI--superconductor system focused on the topological phase diagram
\cite{Gilbert_TI_SC_PRB16},
and the most recent theoretical studies propose gate control of MZMs
in ribbon geometries with large aspect
ratios \cite{ MacDonald_QAH_MF_PRB18,2018_KTLaw_PALee_PRB}.

\begin{figure}[t]
\includegraphics[width=0.4\textwidth]{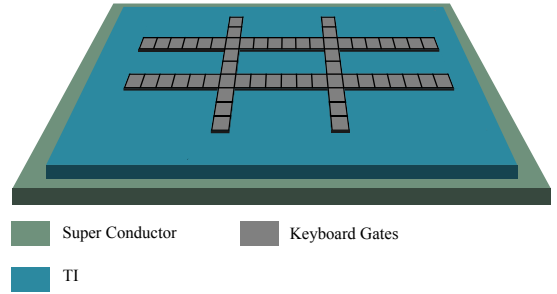}
\caption{A quantum anomalous Hall insulator/ superconductor heterostructure.
The crossbar shaped gates at the top, can change the electro-static
potential of the top surface locally.\label{fig:Device_scheme}}
\end{figure}

In this letter, we build upon that recent work.
We theoretically demonstrate the
micron--scale gate dimensions required for creating MZMs, and we analyze how
gate geometry effects the gate voltage required to create the MZMs.
The system under consideration is an array of `keyboard' gates \cite{MacDonald_QAH_MF_PRB18}
on top of the MTI / superconductor bilayer as illustrated in Fig. \ref{fig:Device_scheme}.
The effect of the geometric shape of the gated area on unwanted
hybridization and the topological band gap is analyzed.
Fundamental building blocks of the crossbar gate-array, the I-shaped and the T-shaped gates,
are demonstrated.
To ensure that the MZMs are not trivial low energy modes,
the symmetry of the MZM wave functions are analyzed to show that the
the wave function is its own complex conjugate.
%
%

\begin{figure*}[t]
\includegraphics[width=1.0\textwidth]{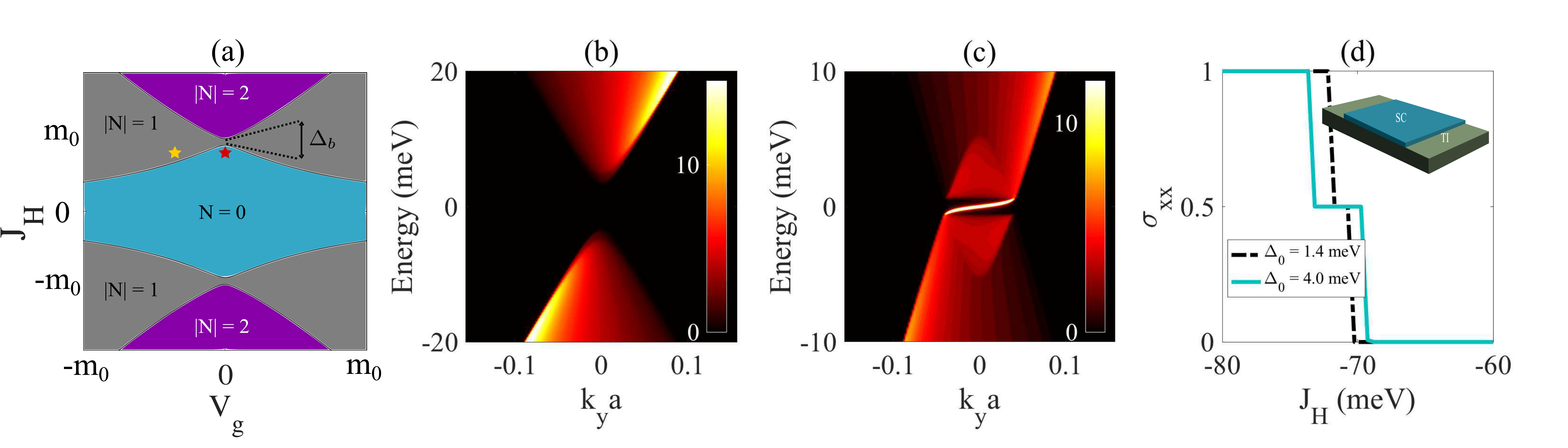}
\caption{
(a) Phase diagram of the system with $\Delta_{t}$ = 0.  $V_{g}$ is applied to the top surface.
(b) and (c) show the spectral function $A(k_y,E)$ at the edge site ($x=0$) of a semi--infinite plane
($-\infty < x \le 0$ and $-\infty < y < \infty$) at different gate voltages.
(b) $V_{g}$ = 0.0 mV corresponding to the red star in (a)
and (c) $V_{g}$ = -20 mV corresponding to the yellow star in (a).
(d) The half-integer plateau in conductivity of a 100 nm wide
by 40 nm long
MTI / superconductor bilayer with topological insulator leads for two different values of
$\Delta_t$ as shown in the legend.
$E_f = 0.1$ meV, $V_g = 0$, and $\Delta_b = 0$.
Inset: Illustration of the structure.
\label{fig:Phase-diagram-transmission}
}
\end{figure*}

The system, as shown in Fig. \ref{fig:Device_scheme},
consists of a thin-film MTI placed on the top of an s-wave superconductor.
Electric gates on top of the MTI control the top-surface electrostatic potential.
The Hamiltonian of the system is \cite{chung_conductance_2011}
%
\begin{equation}
H_{BdG}=\left(\begin{array}{cc}
H_{0}(k)-\mu & \Delta_{k}\\
\Delta_{k}^{\dagger} & -H_{0}^{*}(-k)+\mu ,
\end{array}\right).
\label{eq:HamiltonianMatrix}
\end{equation}
where
\begin{equation}
\Delta_{k}=\left(\begin{array}{cc}
i\Delta_{t}\sigma_{y} & 0\\
0 & i\Delta_{b}\sigma_{y}
\end{array}\right),\label{eq:ParingPotential}
\end{equation}
and
\begin{equation}
H_{0}(k)=\hbar v_{f}(k_{y}\sigma_{x}-k_{x}\sigma_{y})\tilde{\tau_{z}}+m(k)\tilde{\tau_{x}}+J_{H}\sigma_{z}+\frac{V_{g}}{2}(\tilde{\tau_{z}}+I).
\label{eq:TI_Hamiltonian}
\end{equation}
%

$\Psi=[(\psi_{\uparrow t}\ \psi_{\downarrow t}$\ $\psi_{\uparrow b}\ \psi_{\downarrow b}), \ (\psi_{\uparrow t}^{\dagger}\ \psi_{\downarrow t}^{\dagger}\ \psi_{\uparrow b}^{\dagger}\ \psi_{\downarrow b}^{\dagger})]$ is the basis of the Hamiltonian where $\psi_{\uparrow t}(k)$
corresponds to an up-spin electron on the top surface and $\psi_{\downarrow b}^{\dagger}(k)$
corresponds to a down-spin hole at the bottom.
${\bm \sigma}$ and $\tilde{\bm \tau}$ are Pauli matrices corresponding to the spin and the
top-bottom surfaces, respectively.
$\Delta_{t}$ and $\Delta_{b}$ are the proximity-induced Cooper-pairing interactions at the top and
bottom surfaces, respectively.
The pairing interaction of the bottom surface that is in contact with the superconductor is
$\Delta_b = 1.4$ meV \cite{peter_schackert_2015},
and at the top surface, $\Delta_{t} = 0$.
$m(k) = m_0 + m_1k^2$ represent the hybridization of the top and bottom surfaces of the thin--film MTI.
The value of these terms for 3 quintuple layers of Bi$_2$Se$_3$ are
$m_0 = 70$ meV and $m_1 = 18$ eV{\AA}$^2$ \cite{zhang_engineering_2017, zhang_crossover_2010}.
The quantity $\hbar v_{F} = 3.29$ eV{\AA} is consistent with DFT results \cite{liu_model_2010}.
$J_{H}$ is the Hund's rule coupling from the ferromagnetic exchange interaction induced by the Cr dopants.
For calculations using a fixed $J_H$, the value is $J_H = 65$ meV.
The chemical potential $\mu = 0$.
The last term in Eq. (\ref{eq:TI_Hamiltonian}) represents
the gate voltage $V_{g}$ applied at the top surface of the MTI, with the bottom surface adjacent to the
superconductor at ground.
An equivalent approach would be to shift $\mu$ by $-V_g/2$ in Eq. (\ref{eq:HamiltonianMatrix})
and apply the gate voltage symmetrically across the top and bottom layers
such that the bottom layer is shifted to $-V_g/2$ and the top layer is shifted to $+V_g/2$.
This latter approach is the way the gate voltage was included
in recent work \cite{MacDonald_QAH_MF_PRB18}.

\begin{figure}[h]
\centering{}\centering \includegraphics[width=0.52\textwidth]{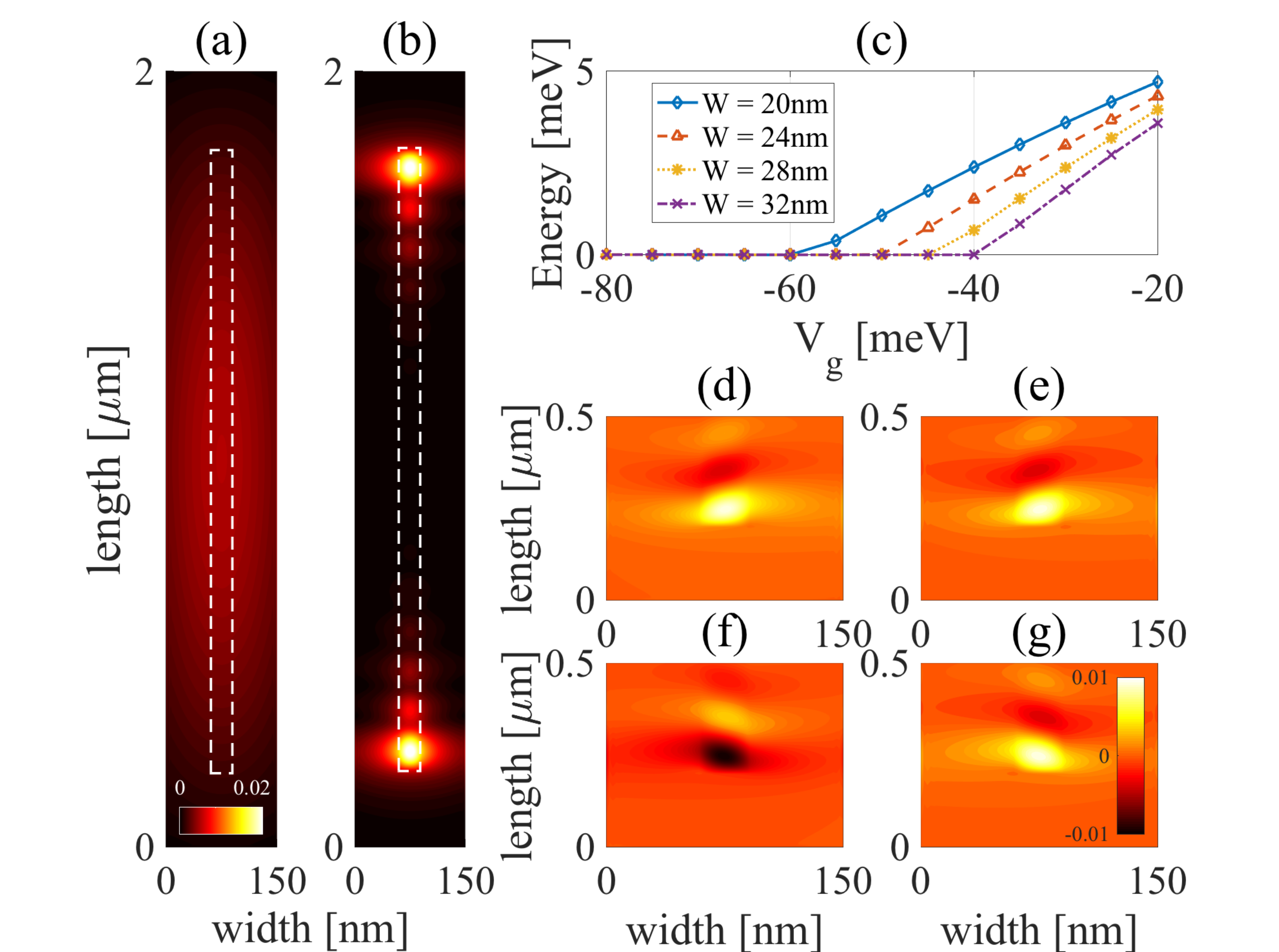}
\caption{
(a) The lowest positive--energy state at $V_{g}=-20$ mV.
It transitions into (b) two MZMs at $V_g = -50$ mV.
(c) Ground state energy as a function of the applied voltage for different widths.
The length is fixed at 1.6 $\mu$m.
Components of the MZM used to verify that the zero--energy mode is indeed a Majorana mode:
(d) $\textrm{Re}[\langle\gamma|\psi_{\uparrow b}\rangle]$
(e)$\textrm{Re}[\langle\gamma|\psi_{\uparrow b}^{\dagger}\rangle]$ (f) $\textrm{Im}[\langle\gamma|\psi_{\uparrow b}\rangle]$
(g) $\textrm{Im}[\langle\gamma|\psi_{\uparrow b}^{\dagger}\rangle]$
}
\label{fig:LDOS}
\end{figure}

\begin{figure}[t]
\centering \includegraphics[width=0.5\textwidth]{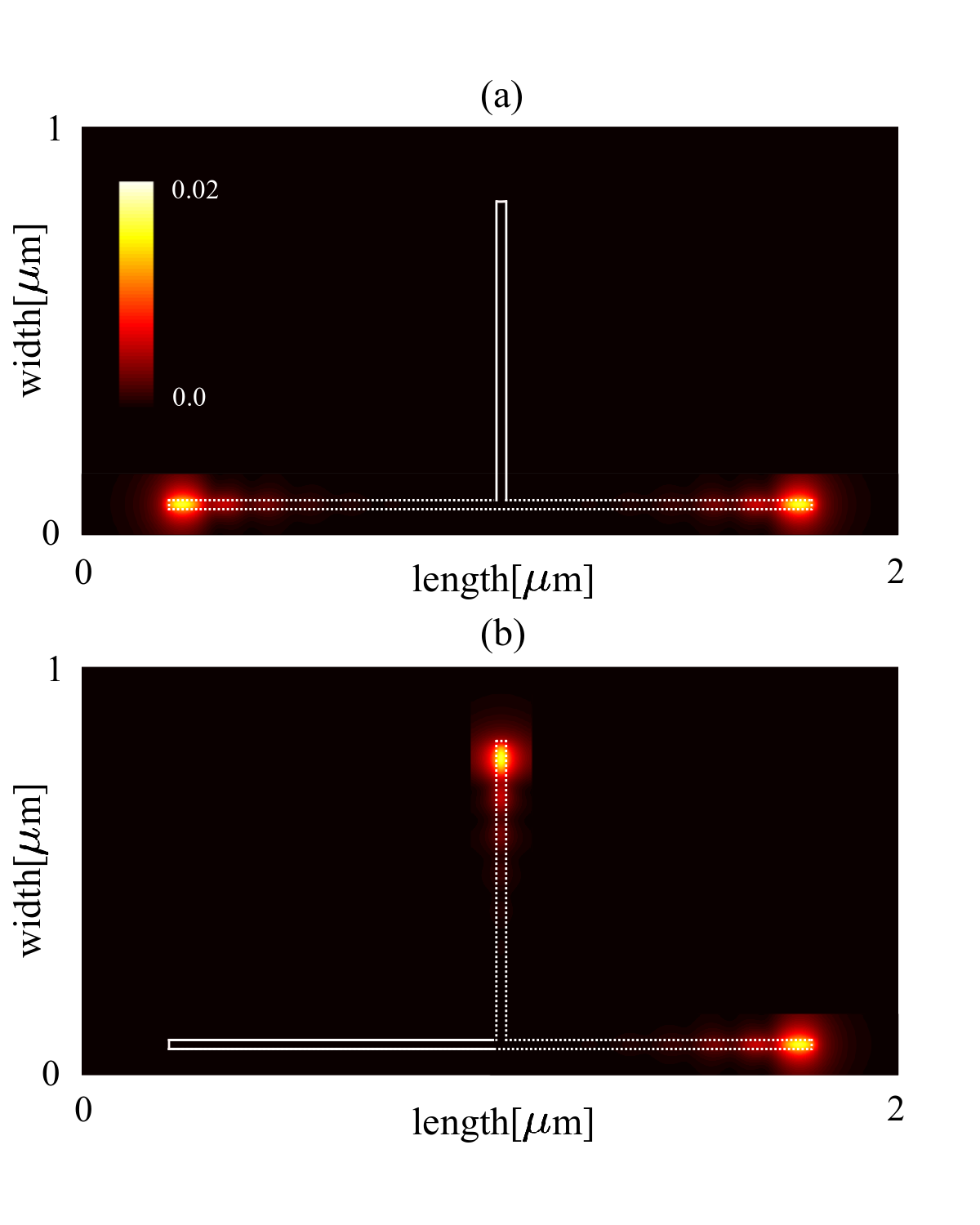}
\caption{
Shifting the MZM in the left (a) to the top (b) by changing the
gate electric potential.
The gates are set to be on and off inside the dashed and solid lines, respectively.
The widths of the gated regions are 28nm.
\label{fig:T_junction}
}
\end{figure}

To model finite and spatially varying structures,
the Hamiltonian is transformed into a tight--binding model on a square lattice by
substituting $k_{x}\rightarrow-i\frac{\partial}{\partial x}$ and
$k_{y}\rightarrow-i\frac{\partial}{\partial y}$
in Eq. (\ref{eq:HamiltonianMatrix}) and discretizing the derivatives
using a 1 nm discretization length \cite{Gershoni_IEEEJQE_93}.
Each site in the tight-binding model is then represented by an $8 \times 8$ matrix corresponding to
Eq. (\ref{eq:HamiltonianMatrix}) with the inter--site matrix elements coming from the discretized derivatives.
Eigenenergies and eigenstates of the discretized, spatially--varying systems
are calculated numerically using a Lanczos algorithm.
The eigenstate calculations use periodic boundary conditions,
and the simulation domain is sufficiently large that the gated regions
in the neighboring cells do not interact with each other.
Simulation domains are illustrated in Figs. \ref{fig:LDOS}(a,b)
and \ref{fig:T_junction}(a,b).

For the calculation of conductance shown in Fig. \ref{fig:Phase-diagram-transmission}(d),
the transmission is determined in the usual way from the `device' Green's function
and the lead self-energies \cite{LakeNemoTheory}.
In the calculation of the lead self energies, an imaginary potential $-i\eta$
with $\eta = 0.1$ meV is placed on the diagonal of the discretized $H_{BdG}$ to assist convergence
of the surface Green's function.
The zero--temperature, two--terminal conductance is then
$\sigma_{xx} = \frac{e^2}{h} T(E_F)$ where $T(E_F)$ is the transmission coefficient at the Fermi energy.

To investigate a single edge mode of a semi-infinite plane
($-\infty < x \le 0$ and $-\infty < y < \infty$)
as shown in Fig. \ref{fig:Phase-diagram-transmission}(b,c),
only the substitution $k_{x}\rightarrow-i\frac{\partial}{\partial x}$ is made, and the derivative is
discretized on the 1 nm grid.
Since the edge of the half-plane is parallel to $\hat{y}$, $k_y$ remains a good quantum number.
The Hamiltonian then becomes a semi-infinite, one-dimensional chain model, where
each site of the chain is represented by a $8\times 8$ $k_y$--dependent
matrix.
The $8 \times 8$ edge Green's function $G^R (k_y, E)$ is calculated
using the decimation method \cite{Sancho:JPF:1985:decimation,Galperin:JCP:2002:decimation}.
Note that this is traditionally referred to as the `surface Green's function,'
however, for this system, the `surface' is an `edge.'
To resolve the edge spectrum, the energy broadening $\eta$ used in the
calculation of the surface Green's function
is 1 meV, which is chosen to be five times larger
than the energy discretization step size.
The spectral function at the edge site is
$A(k_y, E) = -2 \: \textrm{Im} \left\{ {\rm tr} \left[ G^R \left( k_y,E \right) \right] \right\}$.

In the superconducting Nambu space, the topological superconducting Chern number (TSC), $N$,
is allowed to be $-2,-1,0,1,2$
where $N$ characterizes the number of chiral edge modes \cite{wang_chiral_2015}.
One practical way to tune the TSC number is to apply an out-of-plane external
electric field to modify the top-surface electrostatic potential
energy \cite{wang_chiral_2015,MacDonald_QAH_MF_PRB18}.
The topological phase diagram of the system represented
by Eq. (\ref{eq:HamiltonianMatrix}) is plotted in Fig. \ref{fig:Phase-diagram-transmission}(a)
as a function of the top-gate potential $V_{g}$ and the magnitude
of the exchange energy $J_{H}$.
The phase boundaries are obtained by the gap closing in the BdG Hamiltonian (Eq. (\ref{eq:HamiltonianMatrix}))
at $k=0$.
To determine the the TSC number in each region,
we evaluate the number of edge states from the bandstructure calculation of a 150 nm wide
ribbon that is periodic along $x$.
$N$ is the number of the degeneracy of the edge states along one edge.
The ribbon width is chosen to be sufficiently wide such that the hybridization of
the edge states is negligible.
The blue area belongs to the trivial phase ($N=0$) of a normal insulator.
The purple regions correspond to $N=2$, which is topologically equivalent to a non-superconducting
quantum anomalous Hall insulator with Chern number $C=1$.
In the grey areas, $N=1$, and a single Majorana edge mode propagates along the edges.
As shown in Fig. \ref{fig:Phase-diagram-transmission}(a), when $V_{g}$ is zero, $N=1$ only occurs over
a narrow range of exchange potentials.
Therefore, gating the top surface can control the transition between different topological phases.

To demonstrate the voltage-controlled topological transition, we numerically
calculate the edge-state spectrum of the semi-infinite plane at different values of $V_{g}$.
A semi-infinite plane is chosen to ensure that the edge state hybridization is zero since the opposite
edge is at $x=-\infty$.
Choosing the parameters for $N\!=\!0$ and $N\!=\!1$
as shown by the two points in Fig. \ref{fig:Phase-diagram-transmission}(a),
the edge spectral function is plotted versus $k_y$ and $E$ in
Figs. \ref{fig:Phase-diagram-transmission}(b) and (c), respectively.
In Fig. \ref{fig:Phase-diagram-transmission}(b) the applied voltage is zero, $N\!=\!0$, and a trivial gap opens
at the Dirac point.
Applying a  $-20\thinspace\textrm{mV}$ potential
to the top surface, a topological transition occurs, and
a gapless Majorana edge mode appears as shown in Fig. \ref{fig:Phase-diagram-transmission}(c).

For further verification of the model,
we construct a 2-terminal, finite-width device
consisting of a central superconducting / MTI bilayer region
with two non-superconducting, topological insulator leads
mimicking the experimental setup recently reported \cite{wang_chiral_2015}.
The structure is illustrated in the inset of Fig. \ref{fig:Phase-diagram-transmission}(d)
where the length of the superconductor area is 40 nm and the width is 100 nm.
As seen in Fig. \ref{fig:Phase-diagram-transmission}(d),
a half-integer plateau in conductivity appears during
a scan of the Hund's-rule exchange energy $J_{H}$,
which emulates a scan of an externally applied magnetic field.
This plateau is the result of a combination
of normal reflection and Andreev reflection \cite{wang_chiral_2015}.
%
%

We now show that a voltage applied to a gate with a large aspect ratio can create localized Majorana zero
modes at the ends.
Fig. \ref{fig:LDOS} shows the simulation geometry that consists of a long, thin, gated region within a
rectangular supercell.
The dimensions of the gated region are 28 nm $\times$ 1.6 $\mu$m,
and the dimensions of the supercell are  150 nm $\times$ 2 $\mu$m.
Fig. \ref{fig:LDOS}(a) is
a color map of of the lowest positive--energy ($E \ge 0$) state
$|\psi_i|^2$ at each site $i$ at a gate voltage of $V_g = -20$ mV.
The thin width of the gated area, $28\thinspace\textrm{nm}$, is less than
the penetration depth of a Majorana edge mode.
This hybridizes the states on the opposing long edges of the gated region,
so that a gap is opened in the energy spectrum and there is no zero--energy mode
along the edges.
Further decreasing  $V_{g}$ to -50 mV,
a pair of bound states appear at the ends of the
gate as shown in Fig. \ref{fig:LDOS}(b),
and  the energy of these bound states drops 5 to 6 orders
of magnitude from 5 meV to $\sim 10^{-9}$ eV,
suggesting that they are MZMs.
The hybridization of the MZMs
at the ends of the gated regions is negligible
since they are 1.6 $\mu$m apart.

%

The voltage at which the MZMs appear depends on the geometry of the gated region.
Fig. \ref{fig:LDOS}(c) shows a calculation of the ground state energy as a function of the gate voltage for
4 different gate widths.
The gate lengths are fixed at 1.6 $\mu$m.
For each gate width, there is a critical gate voltage at which
the ground-state energy goes to zero.
The magnitude of $V_{g}$ required to achieve the zero-energy
state increases as the gate width decreases.
%
%
%
%
%

To confirm that the localized end-modes are indeed MZMs and not simply very
low-energy states,
the eigenvectors $\Psi$ of the zero-modes are analyzed to determine if they
satisfy the property $\Psi = \Psi^\dagger$.
The eight coefficients of each mode at each site $j$ can
be divided into four groups with each of the groups containing a pair
of coefficients that are complex-conjugate, as shown in Eq.
(\ref{eq:Real_Imag_WaveFunction}).
%
\begin{equation}
\begin{aligned}
\Psi = & (A_1-A_2i)\psi_{\uparrow_{t}}+(A_1+A_2i)\psi_{\uparrow_{t}}^{\dagger}\\
 & +(A_1-A_2i)\psi_{\uparrow_{b}}+(A_1+A_2i)\psi_{\uparrow_{b}}^{\dagger}\\
 & +(-B_1+B_2i)\psi_{\downarrow_{t}}+(-B_1-B_2i)\psi_{\downarrow_{t}}^{\dagger}\\
 & +(B_1-B_2i)\psi_{\downarrow_{b}}+(B_1+B_2i)\psi_{\downarrow_{b}}^{\dagger}
\end{aligned}
\label{eq:Real_Imag_WaveFunction}
\end{equation}
%
$\Psi$ is the wave function of a MZM, and $A_{1,2}$, $B_{1,2}$ are
the site--dependent normalization coefficients.
The real and imaginary parts of
$\left\langle \Psi \right.\left|\psi_{\uparrow b}\right\rangle $
and $\left\langle \Psi \right.\left|\psi_{\uparrow b}^{\dagger}\right\rangle $
are shown in Fig. \ref{fig:LDOS}(d)-(g).
Numerically, $\textrm{Re}[\langle\Psi|\psi_{\uparrow b}\rangle]$
and $\textrm{Re}[\langle\Psi|\psi_{\uparrow b}^{\dagger}\rangle]$
are identical, whereas $\textrm{Im}[\langle\Psi|\psi_{\uparrow b}\rangle]$
and $\textrm{Im}[\langle\Psi|\psi_{\uparrow b}^{\dagger}\rangle]$
have different signs, which satisfies Eq. (\ref{eq:Real_Imag_WaveFunction}).
Similar results are obtained for the other bases.
This confirms that the zero--energy states are MZMs.

%

%

The motivation for an array of crossbar gates is to mimic a 1D network of wires
for gate--controlled transfer and exchange of MZMs.
A fundamental building block of such a network is a T--junction
as shown in Fig. \ref{fig:T_junction}.
With voltage applied to the horizontal section of the gate,
two MZMs are created at the ends of the I-shaped gated area.
Turning off the voltage of the left side gate and applying it to the vertical gate
results in the MZM at the end of the `L'.
The MZM does not appear at the sharp corner of the `L'.
Controlling the voltages of the gates moves
the topological regions ($N=1$) and the associated MZMs.
Such a network of top gates can implement a
pixel-by-pixel control of the geometric shape of the topological region,
such that more complicated braiding operations can be achieved within
this scheme.

All of the calculations presented are for 3 quintuple layers.
In terms of the model Hamiltonian (\ref{eq:TI_Hamiltonian}),
only the interlayer hybridization terms, $m_0$ and $m_1$, change due to layer thickness.
For example, at 5 quintuple layers, their values
become $m_0 = 20.5$ meV and $m_1 = 5$ eV{\AA}$^2$ \cite{zhang_crossover_2010}.
The phase diagram of the topological transitions
shown in Fig. \ref{fig:Phase-diagram-transmission}(a) does not change.
This means that the optimum value for $|J_H|$ is approximately $m_0$,
or, in other words, the spin-splitting due to the magnetic exchange interaction from the Cr dopants
should be close to the hybridization gap induced by the inter-surface coupling of the
top and bottom layers.

For the two dimensional system represented by
Eqs. (\ref{eq:HamiltonianMatrix}) - (\ref{eq:TI_Hamiltonian}) with $\Delta_t=0$,
the energy gap at $\Gamma$ is
$E_{\Gamma} = 2|\ep(V_G) - J_H|$ where
$\ep(V_G) = \frac{1}{\sqrt{2}} \sqrt{2m_0^2 + V_G^2 + \Delta_b^2 -
\sqrt{x}}$ with
$x=V_G^4 + 4m_0^2V_G^2 + 4m_0^2\Delta_b^2 - 2V_G^2\Delta_b^2 + \Delta_b^4$.
%
%
Ignoring the $\Delta_b$ terms since $\Delta_b << J_H,m_0$,
at $V_G=0$, the gap $E_\Gamma \approx 2(m_0 - J_H)$ as seen in
Fig. \ref{fig:Phase-diagram-transmission}(b).
The voltage required to close the gap at $\Gamma$ is
$V_G \approx m_0[1 - \sqrt{2(J_H/m_0)^2 - 1}] \approx 10$ mV,
which scales with $m_0$.
When $V_G$ is applied in a ribbon geometry, the hybridization of the two states
along the long edges of the ribbon
effectively increases the parameter $m_0$, which necessitates larger gate voltages
to drive the initial energy gap to zero, as seen in Fig. \ref{fig:LDOS}(c).

Once the bands invert, they take on the Mexican hat shape,
the energy gap moves away from $\Gamma$, and its value is determined
by the proximity induced Cooper pairing potential, $\Delta_b$.
This is the energy gap in which the edge state resides seen in Fig. \ref{fig:Phase-diagram-transmission}(c),
and it is also the gap in which the MZMs reside.
The MZMs are confined at a topological domain wall with the energy
barriers determined by $\Delta_b$, $m_0$, and $J_H$.
Underneath the gated ribbon, the energy spectrum is gapped by the
hybridization energy of the two edge modes.
Outside of the ribbon in the ungated region, the trivial energy gap $2(m_0 - J_H)$
confines the MZMs.
The energy gaps affect the spatial extent of the MZM wavefunctions.
As they are reduced, the increased tunneling necessitates wider and longer gate regions
and greater separation between the gates.
As $J_H$ is reduced, lower temperatures would be required to
maintain the magnetic ordering.
Thus, thicker films with lower inter-surface hybridization require smaller exchange coupling,
and allow lower voltage operation, but at the cost of lower temperatures and larger areas.

In summary,
a gated MTI / superconductor bilayer provides a platform for 2D spatial control of Majorana zero modes.
The phase diagram of the system shows that a gate voltage can control the topological transition
between the $N=0$ and $N=1$ states.
The voltage of the transition depends on the gate width, and narrower gates require larger voltages.
Relatively long gates are required, approximately 2 $\mu$m, to prevent hybridization of the end modes
and to allow the creation of MZMs at low gate voltages.
The MZM positions can be
controlled by the local gating of the top surface.
This scheme may provide a facile method for implementing quantum gates for topological quantum computing.
\\

\noindent
{\em Acknowledgements:} This work was supported by the National Science Foundation under Award NSF EFRI-1433395
2-DARE: Novel Switching Phenomena in Atomic Heterostructures for Multifunctional Applications and
in part by FAME, one of six centers of STARnet, a Semiconductor Research Corporation
program sponsored by MARCO and DARPA..

\newpage
\bibliographystyle{apsrevNoURL}
\bibliography{MyBibTex}

\end{document}